\documentclass[prl,a4paper,10pt,floatfix,superscriptaddress,amsmath,amsfonts,amssymb,preprintnumbers,citeautoscript]{revtex4}

\AtBeginDocument{\usepackage{booktabs}}               
\makeatletter
\g@addto@macro\bfseries{\boldmath}
\makeatother

\usepackage[varg]{txfonts}
\usepackage{color}
\usepackage{hyperref}
\definecolor{cL}{RGB}{32,145,140}
\hypersetup{colorlinks=true,linkcolor=cL,citecolor=cL,urlcolor=cL} 
\usepackage{multirow}
\usepackage{amsmath}
\usepackage{graphicx}
\usepackage{bm}
\usepackage{natbib}

\begin{document}
This manuscript has been authored by UT-Battelle, LLC, under Contract No.
DE-AC0500OR22725 with the U.S. Department of Energy. The United States
Government retains and the publisher, by accepting the article for publication,
acknowledges that the United States Government retains a non-exclusive, paid-up,
irrevocable, world-wide license to publish or reproduce the published form of this
manuscript, or allow others to do so, for the United States Government purposes.
The Department of Energy will provide public access to these results of federally
sponsored research in accordance with the DOE Public Access Plan (http://energy.gov/
downloads/doe-public-access-plan).
\newpage
%
\title{Intermediate Field Spin(on) Dynamics in $\alpha$-RuCl$_3$}

\author{C.L. Sarkis}
\email{sarkiscl@ornl.gov}
\affiliation{Neutron Scattering Division, Oak Ridge National Laboratory, Oak Ridge, Tennessee, 37831, USA}
\author{K.D. Dixit}
\affiliation{Department of Physics and Astronomy, Purdue University, West Lafayette, Indiana, 47907, USA}
\affiliation{Purdue Quantum Science and Engineering, Purdue University, West Lafayette, Indiana, 47907, USA}
\author{P. Rao}
\affiliation{Technical University of Munich, TUM School of Natural Sciences, Physics Department, 85748 Garching, Germany}
\affiliation{Munich Center for Quantum Science and Technology (MCQST), Schellingstr. 4, 80799 München, Germany}
\author{G. Khundzakishvili}
\affiliation{Department of Physics and Astronomy, Purdue University, West Lafayette, Indiana, 47907, USA}
\author{C. Balz}
\thanks{present address: European Spallation Source ERIC, SE 21100 Lund, Sweden}
\affiliation{Neutron Scattering Division, Oak Ridge National Laboratory, Oak Ridge, Tennessee, 37831, USA}
\affiliation{ISIS Neutron and Muon Source, Science and Technology Facilities Council, Rutherford Appleton Laboratory, Didcot OX11 0QX, United Kingdom}

\author{J-Q. Yan}
\affiliation{Materials Science and Technology Division, Oak Ridge National Laboratory, Oak Ridge, Tennessee 37831, USA}
\author{B. Winn}
\affiliation{Neutron Scattering Division, Oak Ridge National Laboratory, Oak Ridge, Tennessee, 37831, USA}
\author{T.J. Williams}
\affiliation{Neutron Scattering Division, Oak Ridge National Laboratory, Oak Ridge, Tennessee, 37831, USA}
\affiliation{ISIS Neutron and Muon Source, Science and Technology Facilities Council, Rutherford Appleton Laboratory, Didcot OX11 0QX, United Kingdom}
\author{A. Unnikrishnan}
\affiliation{Department of Physics and Astronomy, Purdue University, West Lafayette, Indiana, 47907, USA}
\affiliation{Purdue Quantum Science and Engineering, Purdue University, West Lafayette, Indiana, 47907, USA}
\author{R. Moessner}
\affiliation{Max-Planck-Institut für Physik komplexer Systeme, Dresden, Germany}
\author{D.A. Tennant}
\affiliation{Department of Physics and Astronomy, University of Tennessee, Knoxville, Tennessee 37996, USA }
\affiliation{Department of Materials Science and Engineering, University of Tennessee, Knoxville, Tennessee 37996, USA }
\author{J. Knolle}
\affiliation{Technical University of Munich, TUM School of Natural Sciences, Physics Department, 85748 Garching, Germany}
\affiliation{Munich Center for Quantum Science and Technology (MCQST), Schellingstr. 4, 80799 München, Germany}
\author{S.E. Nagler}
\affiliation{Neutron Scattering Division, Oak Ridge National Laboratory, Oak Ridge, Tennessee, 37831, USA}
\affiliation{Department of Physics and Astronomy, University of Tennessee, Knoxville, Tennessee 37996, USA }
\author{A. Banerjee}
\email{arnabb@purdue.edu}
\affiliation{Department of Physics and Astronomy, Purdue University, West Lafayette, Indiana, 47907, USA}
\affiliation{Purdue Quantum Science and Engineering, Purdue University, West Lafayette, Indiana, 47907, USA}

\date{\today}
%
%
%
%
\begin{abstract}
{\centering CLS and KDD contributed equally as first author \par }
 \hfill \break

We present comprehensive inelastic neutron spectroscopic maps of the magnetic field-induced disordered phase of the Kitaev quantum spin liquid candidate material $\alpha$-RuCl$_3$. For fields along both in-plane high-symmetry directions we observe that the spin excitation spectrum at and above a magnetic field of 8~T is gapped. Excitation modes then sharpen for increasing field but are consistently broader than experimental resolution even at 13.5~T. The out-of-plane dispersion diminishes in the 7-10~T regime, signifying enhanced two-dimensional behavior as the in-plane liquid correlations are established. In this regime, excitations are very broad and largely flat for all accessible energy-momenta, which is kinematically at odds with a magnon-decay picture. By contrast, a continuum of fractionalized excitations naturally yields a broad continuum response, which crucially may be accompanied by sharper modes of bound states of fractionalized excitations. Their damping by the continuum accounts for the observed spectral broadening and field dependence. Our results provide strong evidence for the existence of fractionalized excitations in $\alpha$-RuCl$_3$ in a magnetic field.

\end{abstract}

\maketitle

\textbf{Introduction:}
The Kitaev quantum spin liquid (KQSL) phase of the celebrated S = $\frac{1}{2}$ Kitaev model~\cite{kitaev2006anyons} provides fertile grounds in the search for non-Abelian anyons and their application toward topologically protected qubit platforms~\cite{kitaev2003fault,klocke2024spin}. The Van der Waals honeycomb compound $\alpha$-RuCl$_3$ is a prime candidate for realising the KQSL, supported by considerable experimental~\cite{banerjee2016proximate,do2017majorana,banerjee2017neutron,baek2017evidence,little2017antiferromagnetic,janvsa2018observation,banerjee2018excitations,widmann2019thermodynamic,mai2019polarization,balz2019finite,wulferding2020magnon} and theoretical work~\cite{jackeli2009mott,chaloupka2010kitaev,knolle2014dynamics,Li2021identification}. In $\alpha$-RuCl$_3$, the system orders into a zigzag antiferromagnetic (AFM) long-range ordered state below $T_N$ = 7~K in $R~\bar{3}$ single crystals (Fig. 1a) and arises from symmetry-allowed non-Kitaev terms~\cite{rau2014generic,winter2017models}. Application of an external magnetic field $B_\perp$[$B_{||}$] within the honeycomb plane perpendicular [parallel] to a Ru-Ru bond, the magnetic order transforms from a three-layer (ZZ1), to a six-layer (ZZ2) zigzag order at $B_{\textrm{C1}}$= 6~T [7.3~T] before being fully suppressed at $B_{\textrm{C2}}$ = 7.3~T [7.6~T] to reveal a quantum disordered (QD) state (Fig. 1b).~\cite{banerjee2017neutron,lampen2018anisotropic,balz2019finite,schonemann2020thermal,balz2021field}.  While ZZ1 phase has magnetic Bragg peaks only in symmetry-allowed M$_1$ locations (See SI Fig S1), additional magnetic Bragg peaks appeared at the pseudo-$\Gamma$ point location (0~0~1.5) in the ZZ2 phase.

\begin{figure}
    \centering
    \includegraphics[width=\textwidth]{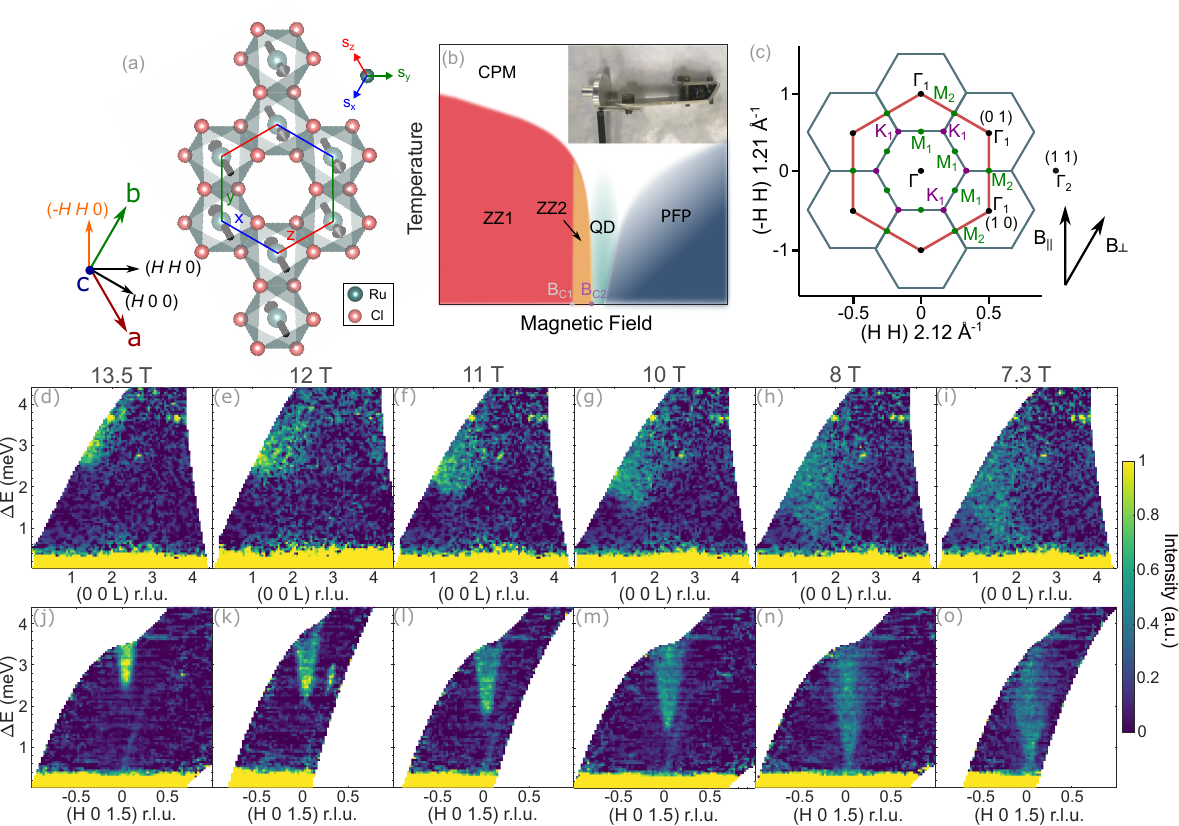}
    \caption{\textbf{Evolution of continuum with $B_\perp$:} (a) Honeycomb layer structure of $\alpha$-RuCl$_3$ showing the bond-dependent Kitaev interactions $K^{x}$, $K^{y}$, and $K^{z}$. We have defined the two applied magnetic field orientations as follows: $B_\perp$ was defined to be along the crystallographic $b$-axis (corresponding to ($\bar{1}~2~0$) in reciprocal space). $B_{||}$ was defined to be along the ($\bar{1}~1~0$) reciprocal space direction. (b) Schematic magnetic field vs temperature phase diagram for $\alpha$-RuCl$_3$ with field applied within honeycomb planes. For zero field at temperatures above the ordering transition the system is a correlated paramagnet (CPM), while below $T_{N}$ it orders into a 3-layer zigzag AFM order (ZZ1). Under an applied magnetic field, a 6-layer zigzag AFM phase (ZZ2) emerges at $B_{\textrm{C1}}$ before entering the quantum disordered (QD) phase at $B_{\textrm{C2}}$. For high fields, the system tends toward a partially field-polarised limit (PFP). (Inset) Picture of 2.0~g sample and aluminum mount. (c) The first (blue honeycomb) and the second (red honeycomb) 2D Brillouin Zones corresponding to panel a with some high symmetry point marked. Under field (directions marked with black arrows) the three zero-field magnetic domains become nonequivalent, as explained in Fig S1. (d-o) INS data taken throughout the intermediate and high-field regime with $B_\perp$. All data were acquired at $T$ = 0.25~K with E$_i$ = 5.5~meV throughout the intermediate and high-field limit (12~T data taken with 6.5~meV, see Methods). (d-i) Out-of-plane momentum (0~0~L) vs energy transfer pseudo-color plots. (j-o) The corresponding in-plane momentum, (H~0~1.5), vs energy transfer pseudo-color plots centered at the local minima (0~0~1.5). Integration ranges for (H~0~0), (-H~2H~0), and (0~0~L) were [-0.076:0.076], [-0.053:0.053], and [1.25:1.75] r.l.u. respectively. Some remnants of the magnet background persist after background subtraction (SI section 1).}
    \label{fig:xtal}
\end{figure}

Previous zero-field inelastic neutron scattering (INS) measurements in $\alpha$-RuCl$_3$  revealed a broad continuum of excitations around the $\Gamma$-point~\cite{banerjee2016proximate,banerjee2017neutron}, which strengthens at 8~T in the quantum disordered phase~\cite{banerjee2018excitations}. The excitations are broad in energy and momentum and repeat every second Brillouin Zone (red hexagon in Fig. 1c) ~\cite{banerjee2017neutron}, indicating dominant nearest-neighbour correlations matching expectations of exact KQSL theory~\cite{baskaran2007exact,knolle2014dynamics,knolle2016dynamics}. Yet the nature of the spectrum in the intermediate field regime ($7 - 13.5$~T) remains considerably debated, with data from different techniques such as Terahertz Spectroscopy (THz)~\cite{little2017antiferromagnetic,wu2018field,shi2018field,reschke2019terahertz,sahasrabudhe2020high}, Raman Spectroscopy~\cite{mai2019polarization,wulferding2020magnon,sandilands2015scattering,zhou2019possible}, INS~\cite{banerjee2017neutron,banerjee2018excitations,balz2019finite,modic2021scale,franke2022thermal}, thermal transport~\cite{kasahara2018majorana,yamashita2020sample,yokoi2021half,czajka2021oscillations,zhang2023sample}, and electron spin resonance (ESR)~\cite{ponomaryov2017unconventional,ponomaryov2020nature, wang2017magnetic}, offering conflicting interpretations. Aside from a topological chiral spin liquid picture comprising of Majorana fermions, the magnetic continuum observed at 8~T has been described in terms of multimagnons broadened by boson decays connected to the field-polarized phase~\cite{winter2016challenges,winter2018probing,ponomaryov2020nature,sahasrabudhe2020high}. Notably, ESR measurements report sharp modes at all fields, which have been interpreted as signatures of magnon modes from long-range spin-ordering~\cite{ponomaryov2017unconventional,ponomaryov2020nature}. 

Here, we utilise the 14~T split coil magnet at the Hybrid Spectrometer (HYSPEC) beamline of the Spallation Neutron Source (SNS) to perform a detailed INS study of $\alpha$-RuCl$_3$ and capture the evolution of the shape of the continuum and its underlying spin gap. The experiments were performed on two (mass $\sim$ 2~g, space-group $R~\bar{3}$, see methods) single crystals of $\alpha$-RuCl$_3$ (Fig. 1c) which show a single N\'eel transition at $T_N$ = 7~K. Field was applied along two distinct crystallographic directions in a 2D layer of $\alpha$-RuCl$_3$: $B_{||}$ and $B_\perp$. Whereas ESR and THz report $q \!\approx\! 0$ resonances, and Raman likewise accesses $(\omega, q \!\approx\! 0)$, INS measures the full $S(q,\omega)$, providing the mode lineshape in both energy and momentum allowing us to track dispersions, damping, and spectral-weight redistribution.

\begin{figure}
    \centering
    \includegraphics[width=\textwidth]{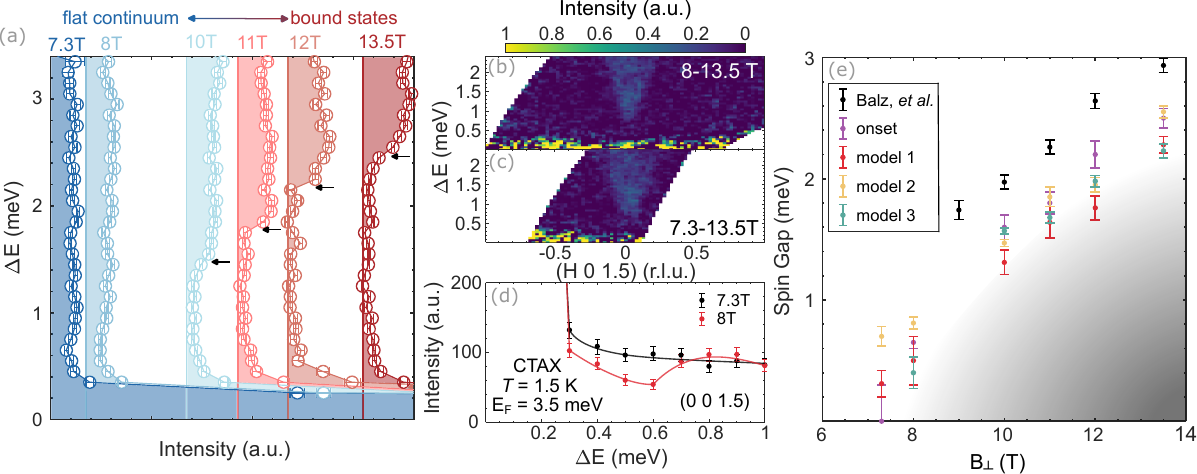}
    \caption{\textbf{Evolution of continuum for $B_\perp$:}  (a) Momentum integrated cuts of data presented in Fig. 1d-o centred on (0~0~1.5) showing a flat continuum at 7-8~T. Intensities are offset linearly in field. Black arrows show onset definitions used in panel e. (b-c) In-plane momentum transfer vs energy transfer pseudocolor plots near (0~0~1.5) for 8~T and 7.3~T respectively, after subtracting the 13.5~T data as a low-energy background. (d) INS data acquired on CTAX for 7.3~T and 8~T. A clear onset is observed in the 8~T data. (e) Evolution of the spin gap at (0~0~1.5) as a function of field applied along $B_\perp$. Model-independent estimates from the data are presented for onset beyond experimental background. The visible first onset (purple dots) of the excitation reflect a more conservative estimate of the spin gap compared to previous INS results (black dots)\cite{balz2019finite}. Fitted estimates using three methods (model 1-3) (red, yellow and green dots) are also provided, particularly for lower field values. Integration ranges in (a-c) for (H~0~0), (-H~2H~0), and (0~0~L) were [-0.076:0.076], [-0.053:0.053] and [1.25:1.75] r.l.u. respectively. For model-specific fits, all errorbars represent a 95\% confidence interval of fitted parameters. }
    \label{fig:INS_H0L}
\end{figure}

\smallskip
\textbf{Gapped continuua for $B_\perp$:} We begin by examining INS data acquired with $B_\perp$ providing data in the scattering plane (H~0~L) plane (equivalent to (-H H L)), using incident energy E$_i$ = 5.5~meV (E$_i$ = 6.5~meV for 12~T, see Methods). Momentum (q) vs energy ($\omega$) transfer pseudocolor plots are presented in Fig. 1d-o starting from $B_\perp$ = 13.5~T and progressively lowered toward the critical field $B_{\textrm{C2}}$ = 7.3~T (see methods section for integration details). For 13.5~T a broad gapped  mode is observed near the edge of our experimental coverage which features a clear dispersion along the out-of-plane (L) direction with energy minima at L = 1.5 in reciprocal lattice units (r.l.u.) and a maximum at L = 3.0, consistent with our previous report~\cite{balz2019finite} and reinforces a three-dimensional picture of the magnetic interactions especially where sharp modes are present. By $B_\perp$ = 11~T the gap at (0~0~1.5) softens and a gapped continuum is revealed. In the 8-10~T regime, the continuum appears remarkably uniform in intensity with a well-defined parabolic dispersive envelope. 
Notably, the out-of-plane dispersion is no longer clearly discernible (Figs. 1h-i) , particularly when compared to the broad bandwidth of the mode. The observed reduction in the effect of interplane interactions suggests a quantum disordered (QD) regime where widespread spin fluctuations and deconfinement diminishes the effect of out-of-plane correlations, as could be expected if a 2D QSL state is established.

The extraordinarily flat, featureless nature of the scattering continuum is quantitatively clear in the momentum integrated cuts at (0~0~1.5) (Fig. 2a), especially at $B_\perp$ = 7.3~T and 8.0~T. As the field increases, a clear gap is visible, which pushes the spectral weights to higher energies. This robust spin gap, clearly discernible at 10~T, monotonically increases with field (Fig. 2a). For 7-8~T, however, a straightforward estimation of the spin gap is challenging due to the low energy background signal arising from incoherent scattering of $^{35}$Cl background. Yet, the seemingly clean $>$2 meV gap allows us to use the 13.5 T dataset as a reasonable proxy for the field-independent background in the low-energy portion of the spectrum for the  7.3~T and 8~T data  (Fig. 2b-c). While 7.3~T still appears ambiguous, the 8~T shows evidence of a clean gap around 0.6~meV. This observation is further validated using the cold-guide triple axis (CTAX) instrument at the High Flux Isotope Reactor (HFIR) using E$_\textrm{f}$ = 3.5~meV on a different $\alpha$-RuCl$_3$ sample (details in Methods). Fig. 2d, taken at (0~0~1.5), shows that the magnetic continuum appears gapless to within instrument resolution, putting an upper bound on the gap at $\Delta_{max}$ $\sim$ 0.25~meV. In distinct contrast, the 8~T the data show a clear minimum followed by an onset near 0.6~meV, consistent with the opening of a gap. The closure of the gap at $B_{\perp}$ = 7.3~T coincides with the field-induced transition into the ordered ZZ2 phase leading to a magnetic Bragg peak at (0~0~1.5)~\cite{balz2021field}.

We quantitatively estimate the spin gap from the low-field data via model-dependent fits. After carefully considering different integration ranges to ensure the appearance of broadening and gap values are unaffected by large integration of the dispersive modes~\cite{do2022gaps} (see SI Sections 3 and 6 for discussion on the integration ranges), we employ three different numerical models (details in Section S2): (M1) a linear onset of intensity assuming a Dirac cone type dispersion; (M2) an empirical fit with a tanh onset at the bottom of the continuum convoluted with instrumental resolution, and finally (M3) the gaped Dirac dispersion relation (Eqn. 8 in Ref. \onlinecite{kitaev2006anyons}) given by $\epsilon(q) \sim \pm \sqrt{((3J^2|q|^2+\Delta ^2))}$ (Fig. S7). The results are presented in Fig. 2e for $B_{\perp}$. These methods provide a more conservative estimate of the spin gap based on the onset of spectral weight, as compared to the resonant energy measurements in NMR and ESR \cite{little2017antiferromagnetic,ponomaryov2017unconventional,ponomaryov2020nature} and triple-axis INS~\cite{balz2019finite}, also because the spin gap is the lowest at the 3D quasi-$\Gamma$-point (0~0~1.5) as compared to any other locations including the 3D zone-center (0~0~0). We can compare with the onset observed directly from the data (black dots). While each method represents a different underlying assumption about the physics that drives the lowest energy excitations’ dispersion, they all consistently provide evidence of the spin gap of at least 0.5~meV at $B_\perp$ = 8~T, and a monotonic, almost linear, increase in the spin gap above 10~T.

 \begin{figure}
    \centering
    \includegraphics[width=\textwidth]{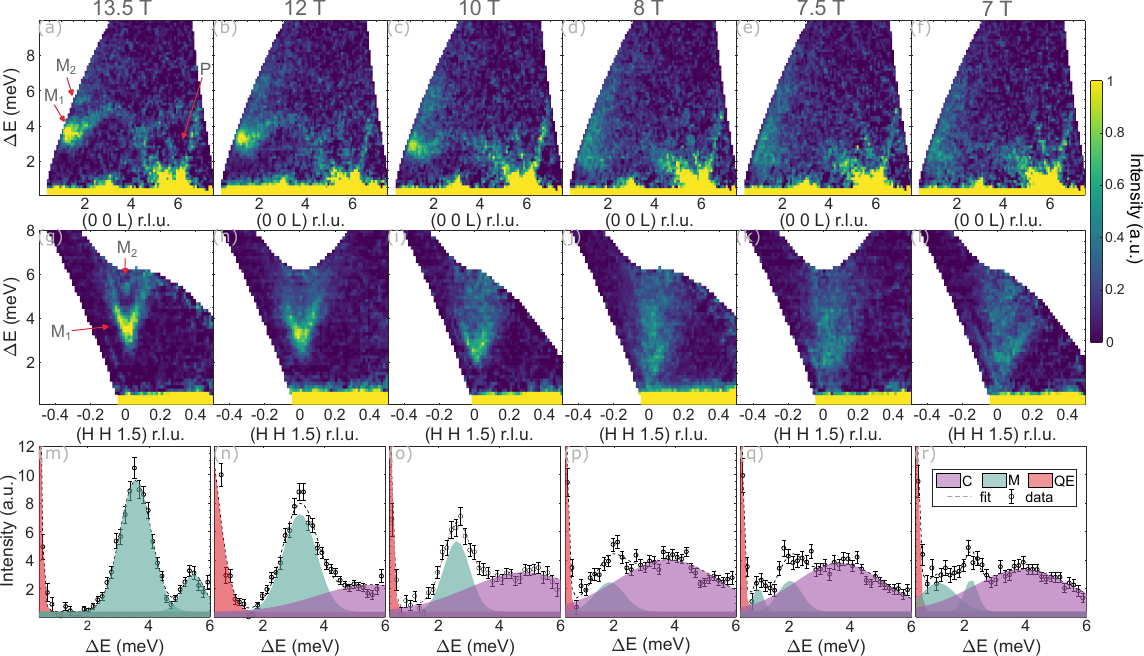}
    \caption{\textbf{Field evolution of spectra for $B_{||}$ between 7~T and 13.5~T:} INS data taken at $T$ = 1.5~K with $E_i$ = 15~meV. (a-f) Out-of-plane momentum (0~0~L) vs energy transfer ($\Delta$E) pseudo-color plots. (g-l) The corresponding in-plane momentum vs energy transfer pseudo-color plots presented at a minimum in the out-of-plane dispersion, (H~H~1.5). Data shows the two sharper modes (M) progressively broaden into continuum-like (C) excitations with decreasing field from 13.5~T to 7~T. Two further features - a sharp feature corresponding to an acoustic phonon mode sharply dispersing from (0~0~6), as well as another spurious feature near (0~0~5) appear to be field independent and hence unlikely to be related to magnetism. (m-r) Momentum integrated cuts centered at (0~0~1.5) as a function of field. For L = 1.5, we observe the smallest observed gap and considerable magnetic intensity. The results for L = 3 is presented in supplementary information with greatly reduced intensity due to the Ru$^{3+}$ form factor~\cite{sarkis2024experimental}. The quasielastic (QE) regime (red) was fit at 13.5~T, where the data was robustly gapped, and held fixed for the rest of the fields. A Gaussian assumption misses the low-energy spin gap (Fig. 4), so the fits are meant to be guides to the eye). As field is decreased, the system undergoes a seemingly smooth collapse of the two sharp modes into a broader continuum. At and below 8~T (near $B_{\textrm{C2}}$=7.8~T), the continuum seems to co-exist with additional sharper modes, likely connected to ZZ2 and ZZ1. Integration ranges for (-H~H~0), (H~H~0), and (0~0~L) were [-0.097:0.097], [-0.043:0.043], and [1.25:1.75] r.l.u. respectively. Errorbars represent 1$\sigma$.}
    \label{fig:INS_15meV}
\end{figure}
\smallskip
\textbf{Melting of modes for $B_{||}$:} We next investigate INS data acquired with field applied along $B_{||}$ where data is collected in the (H~H~L) scattering plane. Fig. 3 shows INS data with $E_i$ = 15~meV and $T$ = 1.5~K. At our maximum measured field of $B_{||}$ = 13.5~T, two sharp features in the data denoted $M_1$ and $M_2$ (Fig. 3a) are observed, consistent with Raman data~\cite{wulferding2020magnon}. Similar to the other field orientation, sharp dispersions are prominent along the out-of-plane direction with energy minima at L = 1.5, 4.5, and maxima at L = 3. The dispersion bandwidth implies interlayer interactions of 0.6~meV at 13.5~T, highlighting the three-dimensional behavior of the system in the high-field limit. The modes appear highly dispersive in-plane with a local minimum at the pseudo-3D $\Gamma$-point, (0~0~1.5). We also observe a sharp acoustic phonon (P) dispersing from the (0~0~6) Bragg peak as well as a bright spurious feature near (0~0~5), which are field independent and are thus unrelated to the magnetism (See SI Section 1 for discussion).

As the field is reduced towards $B_{\textrm{C2}}$, we observe a remarkable broadening of inelastic magnetic features. Already by $B_{||}$ = 12~T, the clarity of M$_2$ is significantly diminished accompanied by the emergence of a broader and more uniform continuum of excitations above M$_1$. At 8~T, closer to $B_{\textrm{C2}}$, even M$_1$ hows a remarkable broadening, accompanied by a progressive softening of the spin gap below M$_1$. Also similar to $B_\perp$ measurements, for $B_{||}$, the out-of-plane dispersion diminishes as the field is reduced, and is almost absent at 8 T compared to the bandwidth, signifying a minimal effect from out-of-plane correlations (Figs. 3d-f) . 

 Momentum integrated cuts centered on (0~0~1.5) provide a quantitative picture (Fig. 3m-r). Cuts confirm well-defined gapped modes at 13.5~T. Their spectral weight shifts to lower energies with field, leaving behind a broader higher-energy continuum. Yet, despite the dominance of a large continuum, sharper features remain in the cuts even below 10 T, in a contrast to $B_\perp$ data. We have phenomenologically fit the cuts in Fig. 2m-r to two Gaussians convolved with instrument resolution . For fields between 8-13.5~T fits retain a relatively sharp lower energy mode (teal peaks), while the higher energy excitation broadens into a progressively dominant continuum (broad purple hump). Below 8~T, we notice additional sharp features. Given that 7.5~T is below $B_{c2}$ = 7.8~T for $B_{||}$~\cite{balz2021field}, these extra modes may be attributed to remnants of the mixed ZZ1 and ZZ2 magnetically ordered phases close to the first-order phase transition, which future tests may confirm.

Cleaner fits were possible at higher fields, where the sharper modes dominate. The FWHM was extracted carefully, by performing cuts with different integration ranges to avoid broadening artefacts from integration of steeply dispersive modes. Fits to momentum integrated cuts of 13.5~T data centered at $L$ = 1.5 and 3.0 show peak centered at 3.53(2) and 4.90(4)~meV, respectively. The extracted full width at half max (FWHM) of M$_1$ was $\delta$E = 1.22(5) at $L$ = 1.5. This FWHM at $L$ = 1.5 is over twice as broad as the elastic instrument resolution ($\delta$E $\sim$ 0.55(5)~meV), although M$_1$ appears to sharpen at $L$ = 3.0 with $\delta$E = 0.63(11). This suggests that, even for fields as high as 13.5~T, the excitations do not comply with conventional magnons, which are generally resolution-limited everywhere. Furthermore, the g-factor extracted from the field evolution of $M_1$ yields g = 4.5(5) (for a single spin-flip $\Delta$S = 1 excitation), which contradict a simple one-magnon picture as well as the published literature values of (g$_{ab}$, g$_c$) = (2.5, 2.3)~\cite{Li2021identification} or (g$_{ab}$, g$_c$) = (2.27, 2.05)~\cite{agrestini2017electronically}. 

\begin{figure}
    \centering
    \includegraphics[width=0.7\textwidth]{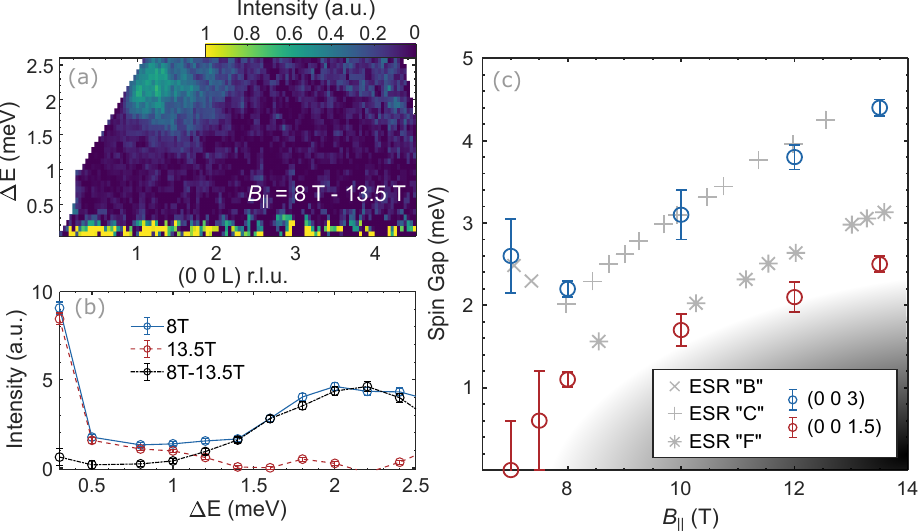}
    \caption{\textbf{Low energy INS data for $B_{||}$:} (a) Out of plane (0 0 L) Momentum vs energy transfer pseudocolor plots of INS data acquired at $E_{i}$ = 6~meV and $T$ = 1.5~K for $B_{||}$. Using the robustly gapped 13.5~T as a subtraction for the 8~T, the low energy portion of the spectra becomes cleanly gapped. (b) Momentum integrated cuts taken at (0~0~1.5) showing the clarity of the gapped onset in the 8~T data.  Integration ranges for (-H~H~0), (H~H~0), and (0~0~L) were [-0.097:0.097], [-0.043:0.043], and [1.25:1.75] r.l.u. respectively.  (c) The Spin gap evolution derived from the onset in Fig. 4b plotted as a function of field for $B_{||}$ for local both the minimum (0~0~1.5) (red circles) and maximum (0~0~3) 3D quasi-$\Gamma$ points (blue circles) (reliably extracted from the data for $B_{||}$~$>$~8~T, while for 7.5~T and 7~T, only upper limits can be estimated.) To compare with ESR measurements reported by Ponomaryov, $et~al.$~\cite{ponomaryov2020nature}, plots of the nearest ESR modes denoted modes ``B'', ``C'', and ``F'' are presented alongside our data.}
    \label{fig:hhl_6meV}
\end{figure}

To zoom-in on the spin gap, we additionally obtain higher resolution data for two fields, $B_{||}$ = 8~T and 13.5~T with $E_i$ = 6~meV, presented in Fig. 4. The data is presented with the 13.5~T data subtracted as a 'background' which helps  remove the spurious low energy features (SI Section 1). A clean gapped mode with an onset $\sim 1.2$~meV (Fig. 4b) is visible.  The low-energy edge of the mode above the gap shows a dispersion along L. This implies that even at $B_{||}=8~T$, a small amount of antiferromagnetic interlayer dispersion survives. 
Together with $E_i$ = 15~meV data we construct the field evolution of the gap for $B_{||}$, shown in Fig. 4c.

\smallskip
\textbf{Spinon origin of the excitations:} The INS response is integrated over non-zero ranges of both energy and momentum, however, the effective energy and momentum resolution does not change with the applied magnetic field. Thus, the observed broadening of the modes in the response near $B_\textrm{C2}$ must be an intrinsic effect of $\alpha$-RuCl$_3$. In some cases, measurements using resonant techniques e.g. ~\cite{wu2018field,sahasrabudhe2020high,ponomaryov2020nature,sahasrabudhe2024chiral}, have been interpreted as showing sharp magnon modes over wide ranges of applied fields. These measurements are sensitive to modes at very long wavelengths, i.e. Q = (0~0~0). Consistent with this, there is reasonable agreement of the observed INS excitation at (0 0 3) (equivalent to (0 0 0)) with ESR modes ``B'' and ``C'' ~\cite{ponomaryov2020nature} (Fig. 4c). These modes are easily distinguishable in ESR and become more well-defined as fields are increased away from $B_\textrm{C2}$. However, the mode ``F'' seen in ESR (grey stars) is much weaker for all fields. The energy of the mode is similar to that of the INS excitation at (0~0~1.5), however as seen in Fig. 3 the INS strengthens and sharpens with increasing field away from $B_\textrm{C2}$. This ``F'' mode has been explained previously as a magnon arising from lowered symmetry along the $c$-axis introduced by DM interactions~\cite{ponomaryov2020nature}. However, stacking faults produce minority regions with different effective periodicities along the $c$-axis, leading to local magnetically ordered phases stable to higher temperatures and critical fields~\cite{bruin2022robustness}. This might explain observed weak modes in resonance experiments close to $B_{\textrm{C2}}$, given their sensitivity to features at wavevectors equivalent to Q = (0~0~0) in the minority regions.

Curiously, Fig. 4 shows that for $B_{||}$ the spectrum at (0~0~1.5) remains gapped at 8~T contrary to the closure of the spin gap reported in heat capacity studies~\cite{tanaka2022thermodynamic,imamura2024majorana}. Our own preliminary heat capacity studies down to $T$ = 1.8~K on our samples seem to corroborate the conclusions of Ref. \onlinecite{tanaka2022thermodynamic} and \onlinecite{imamura2024majorana} (See SI section 8), reflecting that INS probes gapped flux excitations while thermodynamics is sensitive to gapless Majorana modes, as proposed in KQSL scenarios~\cite{knolle2014dynamics}. Also, depending on the microscopic Hamiltonian and the direction of the external field, M-points and K-points can become non-equivalent. Since we can observe only one of the three 2nd BZ M-points (only the one in the scattering plane) and none of the 2nd BZ K-points in our experiment, the possibility of the gap closing at the remaining two out-of-plane M-points or a K-point cannot be fully ruled out~\cite{song2016low}. 

The spin gap observed for both $B_{||}$ and $B_\perp$ appears to increase roughly following a quasi-linear trend with field. A fully linear trend could indicate significant sub-leading $\Gamma$ term contribution in the Hamiltonian, which is expected in $\alpha$-RuCl$_3$~\cite{knolle2014dynamics, winter2017models} although given our coarse field steps, our experiments cannot preclude an unusual non-monotonic behaviour just close to $B_{\textrm{C2}}$ between 7.5 - 9~T, which could constitute a future study.

\begin{figure}
    \centering
    \includegraphics[width=0.7\columnwidth]{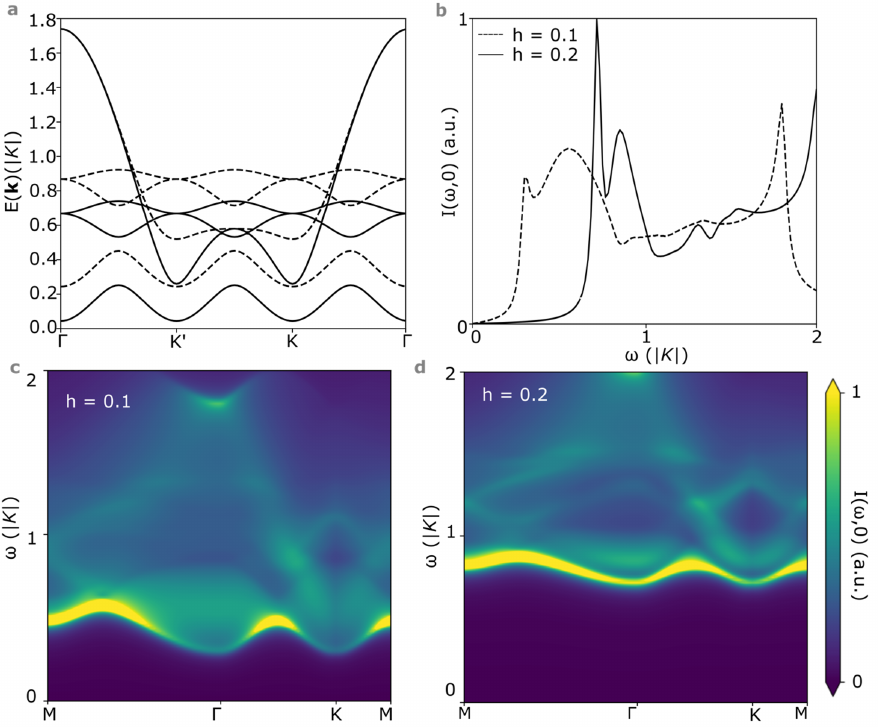}
    \caption{\textbf{Spinon band structure and susceptibility:} (a) Spinon bands at $|B_{||}|$ = 0.1$|K|$ (solid lines) and $|B_{||}|$ = 0.2$|K|$ (dashed lines). (b) Calculated INS intensity at zero momentum for $|B_{||}|$ = 0.1$|K|$ (dashed lines) and $|B_{||}|$ = 0.2$|K|$ (solid lines). As magnetic field increases, the broadened collective mode becomes sharper and shifts to higher energies. (c-d) INS intensity as a function of in-plane momentum and frequency at $|B_{||}|$ = 0.1$|K|$ (panel c) and $|B_{||}|$ = 0.2$|K|$ (panel d). The broad scattering arises from the two-spinon continuum. The two-spinon bound states appear as sharp collective modes, which in turn are broadened by the overlapping continuum.}
    \label{fig:theory}
\end{figure}

An explanation of the broad magnetic excitations based on defects appears unjustified, as random defects would disrupt and broaden all fields uniformly, in contrast to our observations. We observe that the magnetic peak at L = 3.0 for $B_{||}$ is almost resolution-limited (Fig. S12b). Additionally, the phonon mode "P" at (0~0~6) in Figures 3a-f is also resolution-limited, reinforcing the idea that the broadness of the spin modes does not stem from atomic or crystalline disorder. A complicating factor for any attempt to model the experimental findings quantitatively is the absence of an established microscopic model~\cite{maksimov2020rethinking,moller2025rethinking} and the fact that there are no reliable numerical methods for accurately determining the dynamical signatures in the thermodynamic limit~\cite{gohlke2017dynamics}, particularly for bound excitations of a significant spatial extent. Yet, the overall observations help us make considerable conclusions on the nature of the dispersions observed. 

Previously, the observed INS continua~\cite{banerjee2016proximate,banerjee2017neutron,banerjee2018excitations} have been explained with two premises: either taking the presence of a (proximate) QSL as the starting point, with remnants of fractionalized fermionic excitations giving a broad continuum~\cite{knolle2014dynamics,gohlke2017dynamics}; or as an effect of magnon decay from anisotropic interactions of a conventionally ordered magnet~\cite{winter2017breakdown}. Indeed, identifying broad response  as fractionalization isn't straightforward since, even for classical non-linear spin dynamics, the interplay of frustrated interactions and thermal fluctuations may give rise to broad continua in the spin structure factor~\cite{franke2022thermal,bai2019magnetic,zhang2019dynamical}. 

The zig-zag order is destroyed by an applied field, leading to a state with short-range order that can be characterised by a correlation length, $\xi$. In a conventional picture, magnons with wavelengths longer compared to the correlation length cannot propagate, leading to a broad response in the INS spectrum. On the other hand, at short wavelengths, i.e. near zone boundaries where $1/|\mathbf{k}|$ $\leq$ $\xi$  sharp magnon features can be observed~\cite{steiner1976}. Thus, one expects that the longer-wavelength (zone center) magnons broaden more quickly as the field approaches criticality.  The present INS spectra of the field-induced phases of $\alpha$-RuCl$_3$ show an opposite phenomenology: Sharper spin excitation modes persist around (0~0~1.5), which is the 3D zone centre of the ZZ2 phase. Also quite surprisingly, these sharp features at (0~0~1.5) appear when field is moved \textit{away} from $B_{\textrm{C2}}$. Upon lowering the field, these move down in energy and broaden. The continuum too progressively softens, broadens, and appears gapless at the QD $\rightarrow$ ZZ2 transition $B_{\textrm{C2}}$ , where a magnetic Bragg peak is created at (0~0~1.5).  Crucially,  in the 8 T intermediate field regime, the excitations are broad in energy over the entire accessible momenta of the BZ.

While sharp spin flip modes are natural in a field-polarised FM, their broadening via a continuum upon lowering the field is highly unusual. In the conventional magnon-damping scenario~\cite{winter2017breakdown}, a magnon with momentum $\mathbf{k}$ with dispersion $\epsilon_{\mathbf{k}}$ is damped via a continuum of multiple magnon excitations, for example, two-magnons with energy $E^{(2)}_{\mathbf{k}}=\epsilon_{\mathbf{k-q}}+\epsilon_{\mathbf{q}}$. The continuum is made up of the very same magnon excitations, therefore its minimum must be \textit{above} the single magnon dispersion at least for some parts of the BZ (where the minima of $\epsilon_{\mathbf{k}}$ are located). As a result, sharp magnon excitations must exist in at least parts of the BZ and the continuum acquires a distinct momentum dependence. This is in contrast with the behaviour of the observed modes at the pseudo-$\Gamma$ point (0~0~1.5) say at 8 T -- the persistence of only a broad continuum is quite unusual because it is associated with the six-layer ZZ2 Bragg peak below $B_{\textrm{C2}}$ ~\cite{balz2021field}.

Moreover, for this decay picture, it is easy to create a two-magnon excitation at the 3D $\Gamma$ point (0~0~0) where two equal and opposite wavevectors can naturally cancel each other. However, this is not possible at our pseudo-$\Gamma$ point (0~0~1.5). A kinematic analysis reveals that no pair of single-magnon momenta drawn from the ZZ1 or the ZZ2-related magnetic wavevectors alone can sum to (0~0~1.5) (SI Section M). Thus, our observation of a largely flat continuum at this point and throughout the accessible BZ is at odds with the basic (multi)magnon decay scenario. While we cannot exclude yet more complex scenarios of non-perturbative higher-order magnon interactions, we argue in the following that a more natural interpretation arises from the picture of fractionalized spinon excitations consistent with a field-induced KQSL.

The continuum has a distinct field dependence moving up as a function of field. Inter-spinon interactions then lead to broad excitation modes akin to paramagnons of itinerant magnets~\cite{rao2025dynamical}. For increasing fields sharp modes can be pulled below the spinon continuum for $B_{||}$. The sharp spinon bound states carry spin flip quantum numbers and, thus, resemble magnons, but in the presence of a continuum with distinct kinematic restrictions. We argue that the scenario not only explains the observed broad continuum and the broadening of sharp modes, but is also consistent with the behaviour of the low-energy gap.

To that end, we first calculate the spinon band structure and spin susceptibility in a KQSL with magnetic fields along $B_{||}$ $\parallel$ (1~$\bar{1}$~0). We simulate the leading interactions in $\alpha$-RuCl$_3$ by studying the 2D KJ$\Gamma$-model with the 2D spin Hamiltonian:
\begin{equation}
    H = \sum_{<i,j> \in \gamma}\left[ KS_i^\gamma S_j^\gamma + JS_i^\beta S_j^\beta + \Gamma(S_i^\alpha S_j^\beta + S_i^\beta S_j^\alpha ) \right],
\end{equation}

with FM Kitaev coupling and $J$ = 0, $\Gamma$ = 0.8 $|K|$~\cite{maksimov2020rethinking}. The summation is over nearest-neighbour $\gamma$-bonds as shown in Fig. 1a. The computational details are presented in the Methods section and follow the recent Ref.~\cite{rao2025dynamical}. In Fig. 5a, we show the free spinon spectrum from a parton decoupling for $|B_{||}|$ = 0.1$|K|$, 0.2$|K|$, respectively. For the increasing field, the gap increases, and excitations are pushed to higher energies, consistent with the experimental data. 

We next present a calculation accounting for gauge field fluctuations and interactions between spinons in the random phase approximation (RPA)~\cite{rao2025dynamical}.  We focus on the INS intensity defined as:
\begin{equation}
    \mathcal{J}( \omega , k) = \left( \delta_{\alpha \beta} - \frac{k_\alpha k_\beta}{k^2} \right) Im \left[ \chi_{\alpha \beta}(\omega , k) \right],
\end{equation}
where $k_\alpha$ is the projection of the momentum on the spin component directions, and $\chi_{\alpha \beta}(\omega ,k)$ denotes the spin susceptibility tensor. 
Fig. 5b displays the calculated INS intensity $\mathcal{J}$($\omega$,$k$), at zero momentum, which confirms that not only the gap of the structure factor increases but also the intensity distribution sharpens significantly for increasing field, in qualitative agreement with the data. In Fig. 5c-d, we show $\mathcal{J}$($\omega$,$k$) at $|B_{||}|$ = 0.1$|K|$,0.2$|K|$ as a function of in-plane momentum and frequency. We find a rich interplay between the spinon continuum persisting to higher energies and sharpened excitations above the gap with a considerable momentum dispersion. Within our theory, a sharp mode, which is a bound state of two spinons, can appear as a new pole in the RPA equations. For strong binding energy this spinon exciton appears below the edge of the two spinon continuum~\cite{rao2025dynamical} but can also be (over-)damped for weaker interactions between spinons.
We find that for increasing field, the spinon bands are pushed to higher energies and closer together, resulting in a reduced width of the two-spinon continuum. Consequently, the $\textbf{q}$ = 0 mode becomes sharper due to the compressed two-spinon continuum and moves up in energy. In addition, at smaller fields, the continuum is almost gapless at the Gamma point, following the trend of our extracted gap for $\alpha$-RuCl$_3$. Note, in the color map we also see that the mode becomes sharp with a smaller gap away from the $\Gamma$-point. This, however, is outside of the range of our experimental measurements. 

The qualitative agreement of the above theory scenario with the data indicates that a field-induced KQSL displays a continuum of spinon excitations coexisting with bound states of spinons~\cite{rao2025dynamical}. For increasing field, the interaction of continuum and bound states changes, the latter evolving from overdamped to sharp spin flip excitations. We thus can argue that the observed unusual gapped excitations of the intermediate field state are consistent with a KQSL.

Our findings provide a critical constraint on candidate theories for the field-induced quantum spin liquid state in $\alpha$-RuCl$_3$, favouring fractionalized scenarios. Going forward, we envisage a comprehensive set of high-field experiments to determine the Hamiltonian of the material from its high-energy dispersions. Similarly, stringent high-resolution experiments of the low-energy excitations away from the 2D $\Gamma$-point with finer field steps are required, and using high-resolution techniques such as backscattering geometry or neutron spin-echo, involving more sophisticated background treatment (such as subtracting an equivalent chlorine sample to account for the incoherent scattering), to confirm the exact spin-gap throughout the rest of the magnetic Brillouin zone. Finally, rigorous numerical approaches will be crucial for discerning  the spinon-based scenarios for a 3D Hamiltonian treatment with modest out-of-plane interactions that account for the L modulations of the low-energy part of the continuum just above the spin-gap.

\section{Methods} 

\subsection{Crystal growth, quality and specs}
High-quality single crystals of $\alpha$-RuCl$_3$ were grown by sublimating purified reactants following Ref. \cite{yan2023self}. Relatively large (mass$\sim$2~g) single crystals were able to be grown for neutron experiments. We performed the necessary checks on sample quality are presented in the Supplemental Information (SI). The source and synthesis method of the single crystals is an important issue, especially since different synthesis conditions lead to different out-of-plane stackings and slightly different Ru-Cl-Ru angles~\cite{kim2016crystal,imamura2024majorana,yamashita2020sample,zhang2023sample}. While it remains to be seen whether some of the experimental dissimilarities resolve with singly-sourced samples, for this paper we used high-quality single crystal samples grown at 1080$^{\circ}$C using CVT techniques described in Ref. \cite{zhang2023sample} – yielding consistently large crystals in a space group close to $R~\bar{3}$~\cite{mu2022role} space group  confirmed through the systematic absences of the (1~0~0) structural Bragg peak at $T <$ 2~K with lattice constants $a=b=5.975$~\AA, $c=16.93$~\AA, $\alpha=90^{\circ}$, $\beta=90^{\circ}$, $\gamma=120^{\circ}$. 

\subsection{Neutron Scattering}
INS data acquired using the Hybrid-Spectrometer (HYSPEC) at the Spallation Neutron Source located at Oak Ridge National Laboratory used a 14~T symmetric vertical field cryomagnet with the field applied along $\{\bar{1}~2~0\}$ for $B_{\perp}$ and $\{\bar{1}~1~0\}$ for $B_{||}$ (see Fig. 1). At HYSPEC, the detectors are in the horizontal scattering plane and allows a vertical coverage of $\pm$7.5º out of the scattering plane. The crystal was rotated around the vertical (field) axis, leading to a slice of data coverage primarily along (H~0~L) for $B_\perp$ and along (H~H~L) for $B_{||}$ where L represents the out-of-plane reciprocal axis.  

For all (H~0~L) scattering plane measurements, a $^3$He insert read a base temperature of 0.25~K. Data in this orientation were collected using incident energies of E$_i$ = 5.5~meV with a Fermi chopper setting of 240~Hz, leading to an energy resolution FWHM of $\delta$E = 0.18~meV at the elastic line. E$_i$ = 6.5~meV (Fermi Chopper = 360~Hz) was used for 12~T data to crosscheck against spurions and multiple scattering from the magnet. For (H~H~L) measurements the $^4$He cryomagnet used read a base temperature of 1.5~K. Data in this orientation were collected using incident energies of E$_i$ = 15~meV with Fermi chopper speed of 360~Hz and E$_i$ = 6~meV with a Fermi chopper speed of 300~Hz, producing FWHM resolutions of $\delta$E = 0.45~meV and $\delta$E = 0.16~meV at the elastic line, respectively. 12~T data was measured with a $^3$He insert installed, which was pulled shortly after the beginning of the experiment and thus features a slightly different background profile (see supplemental for more details). Typical collection times were 5 hours for E$_i$ = 15~meV data and 14 hours for lower E$_i$ data with samples rotated in 1$^\circ$ steps covering an angular range of 120$^\circ$ centered on (0~0~1.5).

Neutron data were prepared and visualized using Mantid software~\cite{arnold2014mantid}. To avoid artefacts of broadening from integrating steeply dispersing spectra, all integration ranges provided below were determined by testing different integration ranges at the bottom of the parabolic mode at (0~0~1.5) in the 13.5~T data (See SI). HYSPEC features vertical focusing which broadens the intrinsic vertical Q resolution, allowing for a slightly coarser integration in that direction. For all presented pseudocolor slices and momentum integrated cuts integration along (0~0~L) were $\pm$0.25 r.l.u. ($\pm$ 0.09 $\AA^{-1}$). In all $B_\perp$ data integration along (H~0~0) were $\pm$0.076 r.l.u ($\pm$ 0.08 $\AA^{-1}$), with integration orthogonal to the scattering plane along (-H~2H~0) were $\pm$0.053 r.l.u. ($\pm$0.11~$\AA^{-1}$). Finally, in all $B_{||}$ data integration along (H~H~0) were $\pm$0.043 r.l.u ($\pm$ 0.09 $\AA^{-1}$), with integration orthogonal to the scattering plane along (-H~H~0) were $\pm$0.097 r.l.u. ($\pm$0.11~$\AA^{-1}$).

INS data acquired using the cold triple-axis spectrometer (CTAX) at the High-Flux Isotope Reactor located in Oak Ridge National Laboratory used an 8~T asymmetric vertical field cryomagnet aligned such that the field was applied along a \{$\bar{1}~1~0$\} equivalent direction ($B_\perp$).  A fixed final energy of E$_\textrm{f}$ = 3.5~meV was chosen with varying incident energies to reach total energy transfers of $\Delta$E = 1.5~meV. Cooled Beryllium filters were installed on the scattered side and all measurements were taken using collimation settings of 48-40-40-120. This configuration gave an energy resolution of $\delta$E = 0.11~meV at the elastic line. All measurements were taken reading a base temperature of 1.5~K. Typical count times were 30 minutes per point to collect accurate counting statistics. 

To prevent sample movement from magnetic torque, all samples were enclosed in a large aluminum mount (see Fig. 1). All data presented subtract background coming from the empty magnet. Just the subtraction of a magnet background or the so-called “empty-can” fails to remove multiple scattering from the sample and the large $^{35}$Cl incoherent background present in $\alpha$-RuCl$_3$. Instead, where noted we choose to use the low energy portion of 13.5~T data below the observed spin gap as a background for the 7 - 8~T data. The caveat of this procedure are the risks subtracting out quasielastic magnetic excitations (if present) which could tend to overestimate the spin gap. 

\subsection{Heat capacity}
Heat capacity measurements were performed on small single crystals (m = 0.5~mg) using a Quantum Design Dynacool Physical Properties Measurement System (PPMS) at Birck Nanotechnology Center at Purdue University with fields applied along both in plane high-symmetry directions,  $B_\perp$ and $B_{||}$. To confirm absence of torque during the measurement, sample orientations were confirmed prior to and after measurement using a Photonic Science x-ray Laue diffractometer.

\subsection{Theoretical modelling}
To extract the spinon spectrum, we use Kitaev's map from spin to Majorana fermions $S_i^{\alpha} \rightarrow ic_i b_i^{\alpha}/2$. The Hamiltonian is then decoupled within a Majorana mean field theory with the following channels:
\begin{equation}
    \eta_\gamma = i<c_ic_j>_\gamma,\quad Q_\gamma^{\alpha \beta} = i<b_i^\alpha b_j^\beta >_\gamma,\quad m^\alpha = i<c_i b_i^\alpha>.
\end{equation}

Here $<(\ldots )>_\gamma$ denotes averaging over bilinears across the $\gamma$-bond. We incorporate the effect of magnetic field by adding the Majorana bilinears~\cite{ralko2020novel}:
\begin{equation}
    \frac{|B_{||}|}{4} \sum_{i,j\epsilon N.N.N.} ic_ic_j - |B_{||}| \sum_{<i,j> \epsilon \gamma.} ib_i^\gamma b_j^\gamma.
\end{equation}
To simulate the spinon spectrum in Fig. 5a in the main text, the MF parameters are taken to have the values:
\begin{equation}
    \eta_\gamma = 0.52,\quad Q_\alpha^{\alpha \alpha} = -1,\quad Q_\gamma^{\alpha \beta} = 0.1,\quad \textbf{m} = 0.1|\textbf{B}_{||}|(\bar{1}~1~0);\ (\alpha \neq \beta \neq \gamma).
\end{equation}
The spin susceptibility is calculated in the random phase approximation (RPA) by summing the geometric series of one-loop spinon diagrams as pioneered by Anderson and collaborators~\cite{ho2001nature}. For this purpose, we first consider the susceptibility tensor of non-interacting spinons in the Matsubara representation:
\begin{equation}
\begin{aligned}
    \left[\chi_0 (\omega,k)\right]_{i\alpha,j\beta} = lim_{ik_0 \rightarrow \omega+i\delta} \sum_{k,l}\int d\tau P_{i\alpha,j\beta}^{(0)} (\tau;r_k,r_l ) e^{ik_0 \tau-ik.(r_k-r_l ) }\\ 
    P_{i\alpha,j\beta}^{(0)} (\tau;r_k,r_l ) = \langle T_\tau \{ ic_i (\tau,r_k) b_i^\alpha(\tau,r_k )ic_j(0,r_l ) b_j^\beta (0,r_l) \} \rangle_0,
\end{aligned}
\end{equation}

where the $T_\tau$ is time ordering with respect to the imaginary time $\tau$, and $ik_0 \rightarrow \omega+i\delta$ is the analytical continuation. The indices i,j=A,B denote sublattice and $\alpha,\beta$ are spin indices. The details of computing the tensor $\chi_0 (\omega,k)$ are published elsewhere~\cite{rao2025dynamical}. The susceptibility in the RPA approximation is then given by the following ‘matrix equation’:
\begin{equation}
    [\chi(\omega,k)]_{i\alpha,j\beta} = [\chi_0 (\omega,k)]_{i\alpha,k\gamma} [ 1 + 0.6\hat{U}(k) \chi_0 (\omega,k)]_{k\gamma,j\beta}^{-1}.
\end{equation}

The interaction matrix elements $\hat{U}(k)$ is determined by the Hamiltonian and we have included an additional numerical factor 0.6. The spin susceptibility is given by:
\begin{equation}
    \chi_{\alpha \beta} (\omega,k) = \sum_{i,j=A,B} [\chi(\omega,k)]_{i\alpha,j\beta} e^{-ik.(r_i-r_j) }
\end{equation}
where $r_i$ and $r_j$ are sublattice positions inside a unit cell.

\section{Acknowledgements}
The authors thank Gabor Halasz, Radu Coldea and Jason Alicea for helpful suggestions and Melissa Graves-Brook for providing excellent support on the beamline. AB and KD additionally thank Bishnu Belbase for help with sample analysis and alignment. The neutron scattering and crystal growth work at ORNL and UTK is supported by the U.S. Department of Energy, Office of Science, National Quantum Information Science Research Centers, Quantum Science Center. The heat capacity and neutron data analysis at Purdue are supported by the U.S. Department of Energy, Office of Science, Basic Energy Sciences, Grant No. DE-SC0022986. This work was in part supported by the Deutsche Forschungsgemeinschaft under grants SFB 1143 (project-id 247310070) and the cluster of excellence ctd.qmat (EXC 2147, project-id 390858490). JK acknowledges support from the Deutsche Forschungsgemeinschaft (DFG, German Research Foundation) under Germany’s Excellence Strategy (EXC–2111–390814868), and DFG Grants No. KN1254/1-2, KN1254/2-1 TRR 360 – 492547816 [14] and SFB 1143 (project-id 247310070), as well as the Munich Quantum Valley, which is supported by the Bavarian state government with funds from the Hightech Agenda Bayern Plus. JK, PR further acknowledges support from the Imperial-TUM flagship partnership and the Keck foundation. This research used resources at the Spallation Neutron Source and the High Flux Isotope Reactor,  DOE Office of Science User Facilities operated by the Oak Ridge National Laboratory. Beamtime was allocated to HYSPEC on proposal numbers IPTS-27353 and IPTS 29539. Beamtime was allocated to CTAX on proposal number IPTS-29539. Visual representations of the crystal structure shown in this manuscript were produced using the open-source software VESTA~\cite{momma2011vesta}. This manuscript has been authored by UT-Battelle, LLC, under contract DEAC05-00OR22725 with the US Department of Energy (DOE). 

\section{Author Contributions}
AB, CS, KD, SN, and AT conceived the experiment. Single crystals were grown by JQY and were characterised and aligned by CB, KD, CS and AB. KD and CS performed the HYSPEC experiments and CS and GK performed the C-TAX experiments. Heat capacity was performed by AU and GK. The data were analysed by KD and CS with inputs from JK, PR and AU. PR, JK and RM performed the theory in discussions with SE, AT and AB. KD, CS and PR produced the first draft, and all authors contributed to the final draft.

\section{Author Information}
These authors contributed equally as first authors: Kiranmayi Dixit, Colin Sarkis. 

Christian Balz: Present address: European Spallation Source ERIC, SE 21100 Lund, Sweden
\section{Competing Interest Statement}
The authors declare no competing interests.

\end{document}